\documentstyle[aps,prl,epsf]{revtex}
\begin{document}

\draft

\title{Collimated Electron Jets by Intense Laser Beam-Plasma
Surface Interaction under Oblique Incidence}
\author{H. Ruhl\footnote{Hartmut.Ruhl@physik.th-darmstadt.de}}
\address{Theoretische Quantenelektronik, TU Darmstadt, Hochschulstrasse 4A, 
64289 Darmstadt, Germany}
\author{Y. Sentoku\footnote{sentoku@ile.osaka-u.ac.jp}, K. Mima,
K.A. Tanaka$^{(1)}$, and R. Kodama}
\address{Institute of Laser Engineering, Osaka University, 2-6 Yamada-oka, 
Suita Osaka, 565, Japan}
\address{$^{(1)}$Department of Electromagnetic Energy Engineering and 
Institute of Laser Engineering, Osaka University, 2-6 Yamada-oka, 
Suita Osaka, 565, Japan}
\date{\today}
\maketitle

\begin{abstract}
Oblique incidence of a $p$-polarized laser beam on a fully ionized 
plasma with a low density plasma corona is investigated numerically
by Particle-In-Cell and Vlasov simulations in two dimensions. 
A single narrow self-focused current jet of energetic electrons 
is observed to be projected into the corona nearly normal to the
target. Magnetic fields enhance the penetration depth of the electrons
into the corona. A scaling law for the angle of the ejected electrons 
with incident laser intensity is given.
\end{abstract}

\newpage

The availability of tabletop high intensity laser systems has
lead to the investigation of novel regimes of short pulse
laser-plasma interaction. Recently the emission of collimated 
electron jets under specular angles with respect to the density 
normal direction have been observed for an obliquely incident 
laser beam on a steep density plasma \cite{bastiani}. 

When a target is irradiated by an intense laser pulse above
the field ionization threshold it quickly ionizes \cite{bauer}. 
For sufficiently long laser pulse irradiation the plasma present 
on the surface gradually expands into the vacuum with the ion 
acoustic speed. Hence, a plasma corona is formed. For short 
laser pulses there is not enough time for hydrodynamic
expansion. Short pulse simulations show however, that an ion shelf 
is formed on a typical time scale $t_{\text{s}}=\omega_{\text{0}}^{-1}
(m_{\text{i}}/Zm_{\text{e}})^{1/2}$ due to the generation of a
strong electric field at the plasma-vacuum interface \cite{gibbon0}.
This ion shelf represents a low density plasma corona.

There are different mechanisms which can lead to collimated
electron jets when an intense laser pulse interacts with a
vastly over-dense, steep density plasma that has a low density 
plasma corona. One effect that plays a role in the interaction 
is the Brunel effect \cite{brunel} which works for oblique incidence. 
Here the electrons are accelerated into the vacuum as well as 
into the target by the electric field present along the density 
gradient. The coronal plasma is expected to collimate and enhance
the range of the ejected electrons in two (2D) or three (3D) spatial 
dimensions by quasi-steady magnetic field generation. The collimating 
effect of quasi-steady fields has recently been addressed in 
\cite{gorbunov,pukhov} in a different context. 

To investigate the phenomenon just outlined in more detail
we perform Particle-In-Cell (PIC) and Vlasov (VL) simulations
both in two spatial dimensions (2D). In both approaches we 
do not simulate the evolution of the corona self-consistently
but treat it parametrically with ions fixed. In our PIC simulations
we rotate the target and in our VL simulations we boost to
the frame of normal incidence to model oblique incidence of 
the laser beam. The boost frame method is well established 
in 1D \cite{bourdier,ruhl1,gibbon1}. However, in 2D using the 
boost frame is very helpful in establishing the physics of the
underlying laser-plasma interaction.

We investigate the interaction of a $p$-polarized laser beam 
incident under angles of $30^{\circ}$ (PIC) and $45^{\circ}$
(VL) on a preformed fully ionized target with an under-dense 
plasma corona in front of it. In both simulations the laser beam has a
duration of about $100 \text{fs}$. For the PIC case the laser beam intensity 
is $2.0 \cdot 10^{18} \text{W/cm}^{2}$ and for the VL case $1.0 \cdot
10^{17} \text{W/cm}^{2}$. The laser wavelength in the simulations
is $1 \mu \text{m}$ with beam diameters of $8 \; \mu \text{m}$ (PIC)
and $5 \; \mu \text{m}$ (VL) at full-width-half-maximum. The 
coordinates of the simulation box are $x$ and $y$, respectively. 
The size of the simulation box is $23 \; \mu \text{m} \times 23 \; \mu 
\text{m}$ for the PIC simulations and $6 \; \mu \text{m} \times 12 \;
\mu \text{m}$ for the VL simulations. In our PIC simulations we
assume mirror-reflecting boundary conditions for the electrons in 
$x$- and $y$-directions. In the VL simulations we use periodic 
boundary conditions in $y$-direction. Electrons leaving the simulation 
box at $x=6\mu \text{m}$ are replaced by thermal electrons and at 
$x=0 \mu \text{m}$ they are allowed to escape. We note that mirror-reflecting 
boundary conditions in our PIC simulations force us to increase the 
simulation box for long simulation times. 
The distribution functions for the electrons and ions needed for 
the VL simulations have two momentum directions $p_{\text{x}}$ 
and $p_{\text{y}}$ in addition. The quasi-particle number per cell 
used in the PIC simulations is $50$ for each species. The fully 
ionized plasma density is $4 n_{\text{c}}$ (PIC) and $8 n_{\text{c}}$ 
(VL). In both simulations  we assume a uniform low density plasma 
corona with a density of $0.1 n_{\text{c}}$ in front of the target.

Plot (a) of Figure \ref{fig:field1} gives the quasi-steady magnetic
field $B_{\text{z}}$ in front of the target inclined at $30^{\circ}$
, as obtained by PIC simulations. The peak magnitude of the normalized 
magnetic field is $0.62$, which corresponds to approximately $30 \; 
\text{MG}$. It changes polarity very rapidly along the density gradient
revealing the presence of a very localized self-focused current jet.
The low density plasma corona guarantees quasi-neutrality 
and helps to generate the magnetic field in front of the target. Plot 
(b) of the same figure shows the electron energy density. We find a collimated 
electron jet which coincides with the quasi-steady magnetic field from 
plot (a). For the parameters of plot (b) the ejection angle is
approximately $17^{\circ}$ form the target normal. There are also fast 
electrons injected into the over-dense plasma. We again observe that 
they are almost normal to the target surface. Figure \ref{fig:field1} 
(c) shows the instantaneous plot of the electron energy density with 
over-plotted positive $B_{\text{z}}$ field indicating the phase of the laser 
field. It is clearly seen that the outgoing electrons are generated on 
the target surface once per laser cycle by the Brunel absorption
mechanism \cite{brunel,ruhl1} and are bunched on the scale of the
laser wavelength consequently. The range of the electrons is enhanced.
A similar result we obtain from our VL simulations which make use of 
boost frame coordinates.

To illustrate how the boost frame approach for oblique incidence
in 2D works we briefly derive the correct boundary conditions for
the laser pulse in the boosted frame. We start by defining an
arbitrary pulse envelope function $z(x,y,t)$ in the lab-frame.
Next we perform a Lorentz rotation of electromagnetic fields
about $(x_{\text{0}},y_{\text{0}})$. In the final step we boost 
the latter to the frame of normal incidence for which the 
longitudinal field $E_{\text{x}}$ disappears. We obtain

\begin{eqnarray}
\label{boost_ep}
E^{B}_x &=& 0 \; , \qquad
E^{B}_y = \frac{1}{\bar{\gamma}} \; 
z \left( x_r,y_r,t \right) \; , \qquad
B^{B}_z  = \frac{1}{c\bar{\gamma}} \; 
z \left( x_r,y_r,t \right) \; , 
\end{eqnarray}

where

\begin{eqnarray}
\label{rotation_x}
x_r&=&\frac{1}{\bar \gamma} (x-x_0) + (y-y_0) \bar \beta \; , \qquad
y_r=\frac{1}{\bar \gamma} (y-y_0) - (x-x_0) \bar \beta \; ,
\end{eqnarray}

with

\begin{eqnarray}
\label{boost_trafo_ct}
t &=& \frac{\bar{\gamma} \bar{\beta}}{c} y^B \; , \qquad
x = -ct^B \; , \qquad
y = \bar{\gamma} y^B \; . 
\end{eqnarray}

The function $z$ is the same function as in the lab-frame.
For the relativistic factors we have $\bar \beta=\sin \theta$ 
and $\bar \gamma = 1/\cos \theta$, where $\theta$ is the angle 
of incidence. Plot (a) of Figure \ref{fig:field2} illustrates
the incident time resolved electromagnetic field $E_{\text{y}}$
for a Gaussian pulse envelope. Plot (b) of the same figure gives
the incident time resolved electromagnetic field $E_{\text{y}}$
of the simulations.

Plot (a) of Figure \ref{fig:field3} gives the quasi-steady magnetic 
field in the plasma corona in front of the over-dense plasma target. 
Plot (b) of the same figure gives the quasi-steady magnetic field 
with the quasi-steady $B^2_{\text{z}}$ over-plotted (red solid lines).
Plot (c) of Figure \ref{fig:field3} gives the quasi-steady magnetic 
field with the quasi-steady longitudinal current density $j_{\text{xe}}$ 
over-plotted (red dashed lines). 

Since the current density $j_{\text{xe}}$ is invariant under Lorentz 
boosts along $y$ it may serve as a quantity from which to determine 
the direction of the electron jets. We now introduce the coordinates 
$\chi =x^{\text{B}}$ and $\xi =y^{\text{B}}+\bar \beta 
c t^{\text{B}}$ which move along with the background plasma current 
present in the boosted frame. Since the time-averaged current density 
$\left< j^{\text{B}}_{\text{xe}} \right>$ in the co-moving coordinates 
varies slowly with time we obtain $\left< j^{B}_{xe} \right> \left( 
x^{B}, y^{B},t^{B} \right) = \left< j^{B}_{xe} \right> \left( \chi, 
\xi \right)$. This yields

\begin{eqnarray}
\label{trafo}
\left< j^{L}_{xe} \right> \left( \chi, \xi \right) = \left<
j^{B}_{xe} \right> \left( \chi, \bar{\gamma} \xi \right) \; .
\end{eqnarray}

The direction of the collimated electron jets in the lab frame can
now be calculated from the direction of the current density in the
boosted frame. Plot (c) of Figure \ref{fig:field3} gives $\left<
j^{B}_{xe} \right>$. The direction of the emitted electrons is
indicated by the white solid line plotted in the figure. We 
obtain a mean emission angle of $20^{\circ}$ in the boosted frame 
and $14^{\circ}$ in the lab frame. We note that the lab frame 
is dilated in transverse direction when viewed from the boosted 
frame and hence the emission angle in the boost frame is larger 
by a factor of $\bar \gamma =1/\sqrt{1-\bar \beta^2}$ as indicated 
by Equation (\ref{trafo}).

In boost frame coordinates we may easily analyze the 
physical mechanism that leads to the large areal quasi-steady 
magnetic field and the direction of the ejected electrons. We 
recall that in the boosted frame we have a constant background 
fluid velocity $u_{\text{B}}=c \sin \theta$ which approaches 
speed of light for large angles of incidence. In this frame the 
polarization of the magnetic field vector of the incident laser 
beam is normal to the $xy$-plane and to the flow direction of the 
background current. If the laser intensity is small enough as
in \cite{bastiani} and the angle of incidence sufficiently large
the boost velocity exceeds the laser quiver velocity. The driving 
force under these conditions is exerted predominantly by the 
oscillating magnetic field of the laser beam (see the red solid
contour lines of $B^2_{\text{z}}$ plotted over $B_{\text{z}}$ in
plot (b) of Figure \ref{fig:field3} for the location of the force). 
The resulting force is ${\bf F}=-e \; {\bf u}_{\text{B}} \times {\bf B}$
and is capable of ejecting electrons out of the surface at a rate 
of once per laser cycle. This is the Brunel mechanism \cite{brunel}. 
The quasi-steady magnetic field in the plasma corona is generated 
by the electron current emitted from the target. The polarization 
of the magnetic field is such that it collimates the electrons 
propagating through the plasma corona.

To derive an approximate criterion for the angle range under which 
the fast electrons are emitted from the target surface we assume 
that the laser target interaction in the boosted frame is 
quasi-one-dimensional. Since the full-width-half-maximum of
the laser beams in our simulations is at least $5 \; \mu \text{m}$
and the intensities are sufficiently low to prevent target imprinting
we believe that this assumption is justified. We next rewrite the 
Vlasov equation in the boosted frame \cite{ruhl3} and 
solve it for an initial Maxwellian. We approximate the plasma-vacuum 
interface by a step-like density profile with $n(x)=n_{\text{0}}$ for 
$x>0$ and treat the ions as immobile. We obtain for the distribution 
function

\begin{eqnarray}
\label{df}
f(t)&=& \frac{n_0}{\sqrt{2\pi}^3 m^3 v^3_{th}} \; 
\exp \left( -\frac{p_x^2(0)+p_z^2(0)}{2m^2 v_{th}^2} \right) \;
\exp \left( -\frac{(p_y(0)+\bar{\beta} \bar{\gamma} mc)^2}
{2\bar{\gamma}^2 m^2 v_{th}^2} \right) \; , 
\end{eqnarray}

and for the equations of motion

\begin{eqnarray}
\label{x}
x(\tau)&=&x-\int^t_{\tau} d\eta \; v_x(\eta) \; , \\
\label{px}
p_x(\tau)&=&p_x+e\int^t_{\tau} d\eta \; \left[ E_x(x(\eta),\eta)+v_y(\eta)
\; \partial_x A_y(x(\eta),\eta) \right] \; , \\
\label{py}
p_y(\tau)&=&p_y+e\left[ A_y(x(\tau),\tau)-A_y(x,t) \right] \; , \\
\label{trajectories_pz}
\label{pz}
p_z(\tau)&=&p_z \; , 
\end{eqnarray}

with

\begin{eqnarray}
\label{vx}
v_{x/y}(\tau)&=&\frac{cp_x(\tau)}
{\sqrt{m^2c^2+p^2_x(\tau)+p^2_y(\tau)+p^2_z(\tau)}} \; . 
\end{eqnarray}

Equations (\ref{py}) and (\ref{pz}) indicate lateral canonical momentum 
conservation in boost frame coordinates. We now assume that $A_{\text{y}}$ 
has a harmonic time dependence. Making use of Equations (\ref{df}) and 
(\ref{py}) and assuming $v_{\text{x}} \ll c$ or $v_{\text{x}} \approx c$
we obtain $\langle p_{\text{y}} \rangle \approx -\bar{\beta} \bar{\gamma}
mc$. The quantity $\langle p_{\text{y}} \rangle$ denotes the ensemble and 
time averaged transverse momentum. Treating $\langle p_{\text{x}} \rangle$ 
as a free parameter and transforming back to the lab frame yields

\begin{eqnarray}
\label{angle}
\langle p^L_y \rangle&=&\bar{\gamma}^2 \bar{\beta}mc
\left( \sqrt{1+\frac{\langle p^2_x \rangle}
{\bar{\gamma}^2m^2c^2}} -1 \right) \; , \qquad
\langle p^L_x \rangle = \langle p_x \rangle \; .
\end{eqnarray}

The ejection angle is now given by $\tan \theta^{'}=
\langle p^L_{\text{y}} \rangle/\langle p^L_{\text{x}} \rangle$.
For $\langle p_x \rangle \rightarrow \infty$ we obtain $\tan \theta^{'} 
= \bar{\beta}\bar{\gamma} = \tan \theta$. This means that only 
ultrarelativistic electrons are ejected at very close to specular 
direction. For smaller longitudinal momenta $\langle p_x \rangle$ we 
expect that the electrons are emitted at angles that are smaller 
than the angle for specular emission as observed in our simulations. 
Assuming that the mean fast electron momentum in $x$-direction
is given by $\langle p_x \rangle / \bar{\gamma} mc \approx 
\sqrt{ \alpha \; I\lambda^2}$ we thus obtain

\begin{eqnarray}
\label{angle1}
\tan \theta^{'}&=&\frac{\sqrt{1+\alpha I\lambda^2}-1}{\sqrt{\alpha
I\lambda^2}} \; \tan \theta \; .
\end{eqnarray}

Equation (\ref{angle1}) looses validity as soon as target deformations
start to become significant. The validity also depends on the accuracy
of the mean longitudinal momentum given as a function of intensity. For
$I\lambda^2=1.0 \cdot 10^{17} \mbox{Wcm}^{-2} \mu \text{m}^2$ we obtain an
ejection angle of $\theta^{'}=14^{\circ}$ and for $I\lambda^2=2.0 \cdot 
10^{18} \mbox{Wcm}^{-2} \mu \text{m}^2$ we obtain $\theta^{'}=17^{\circ}$ 
from the simulations. This yields $\alpha^{-1} \approx 
8.0 \cdot 10^{17} \text{Wcm}^{-2} \mu \text{m}^2$.

In conclusion, we have demonstrated with the help of two different 
simulation techniques that collimated electrons with enhanced range
can be emitted from an over-dense target if a low density plasma corona
is present. In addition, we have shown that fast electrons are injected 
into the over-dense plasma. Both, the ejection and injection directions 
are almost along the density normal direction for $p$-polarized light.
By a transformation to the moving frame in which the laser pulse 
appears to be normally incident we were able to give a criterion 
for the angle range of the emitted electrons with ejection momentum.
We find that for a planar interaction interface only speed of light 
electrons can be emitted at specular direction for $p$-polarized 
light. Less energetic electrons appear under almost normal emission 
angles due to a lack of lateral momentum transfer. This analytical 
result is in qualitative agreement with our numerical observations. 
We note that in addition to the mechanism outlined in this paper other 
mechanisms of fast electron generation like wake-field acceleration 
in the corona may exist leading to different emission angles.

\begin{figure}
\caption[]{Quasi-steady magnetic field $B_{\text{z}}$ (a), 
quasi-steady electron energy density $\epsilon$ (b), and 
instantaneous electron energy density with the laser field 
$B_z$ (c). Yellow contour areas are positive and blue areas 
negative. The parameters are $n/n_{\text{crit}}=4$, $I\lambda^2
=2.0 \cdot 10^{18} \text{Wcm}^{-2} \mu \text{m}^2$, $\theta=
30^{\circ}$, $t=120 \text{fs}$, and $B_{\text{z0}}=100 \text{MG}$.} 
\label{fig:field1}
\end{figure}

\begin{figure}
\caption[]{Illustration of the boost technique in 2D for an
incident laser pulse of Gaussian pulse envelope (a) and time 
resolved $E_{\text{y}}$ taken in the simulations (b). The left
figure in (a) gives the incident pulse, the figure in the middle
gives the pulse after a rotation, and the right figure shows
the pulse after the final boost. The arrows in (a) indicate 
the propagation direction of the laser pulse. The white solid 
lines in (b) give the density profile. The parameters for 
(b) are $n/n_{\text{crit}}=8$, $I\lambda^2=1.0 \cdot 10^{17} 
\text{Wcm}^{-2} \mu \text{m}^2$, $\theta=45^{\circ}$, and 
$t=25 \text{fs}$.}
\label{fig:field2}
\end{figure}

\begin{figure}
\caption[]{Quasi-steady $B_{\text{z}}$ (a), quasi-steady $B_{\text{z}}$ 
with quasi-steady $B^2_{\text{z}}$ over-plotted (b), and quasi-steady 
$B_{\text{z}}$ with current density over-plotted (c). Yellow contour 
areas are positive and blue areas negative. The white solid lines along 
$y$ located between $x=4 \mu \text{m}$ and $x=5 \mu \text{m}$ are density 
iso-contours. The other lines over-plotted in (a) indicate the 
quasi-steady magnetic field. They are along $x=3.63 \mu \text{m}$ 
(solid) and $y=7.18 \mu \text{m}$ (dashed). The parameters are 
$n/n_{\text{crit}}=8$, $I\lambda^2=1.0 \cdot 10^{17} \text{Wcm}^{-2} 
\mu \text{m}^2$, $\theta=45^{\circ}$, $t=75 \text{fs}$, and 
$B_{\text{0}}=1.5 \text{MG}$.}
\label{fig:field3}
\end{figure}

\end{document}